\documentclass[aps,prb,amsmath, superscriptaddress, 10pt]{revtex4-2}
\usepackage{graphicx}
\usepackage{dcolumn}
\usepackage{bm}
\usepackage{mathrsfs}
\usepackage{subfigure}
\usepackage{natbib}
\usepackage{amssymb}
\usepackage{amsfonts}
\usepackage{multirow}
\usepackage{float}
\usepackage{color}
\usepackage{lipsum}
\usepackage{bbm}
\usepackage{caption}
\usepackage{hyperref}
\usepackage{latexsym}
\usepackage{upgreek}
\usepackage{amsmath}
\usepackage{ulem}

\DeclareMathAlphabet\mathbfcal{OMS}{cmsy}{b}{n}

\newcommand{\fbb}{\mathbbmss{f}}
\newcommand{\cbb}{\mathbbmss{c}}
\newcommand{\Mag}{\mathbfcal{M}}

\newcommand{\vecA}{\mathbfcal{A}}
\newcommand{\vecB}{\mathbfcal{B}}
\newcommand{\vecD}{\mathbfcal{D}}

\newcommand{\gbb}{\mathbbmss{g}}

\usepackage{changes}

  \colorlet{Changes@Color}{blue}

\begin{document}

\title{The Type-II/Type-I Crossover in Dirty Ferromagnetic Superconductors}

\author{P. M. Marychev}
\affiliation{HSE University, 101000 Moscow, Russia}
\author{Yajiang Chen}
\email[]{yjchen@zstu.edu.cn}
\affiliation{Key Laboratory of Optical Field Manipulation of Zhejiang Province, Department of Physics, Zhejiang Sci-Tech University, Hangzhou 310018, China}

\begin{abstract}
In this work we investigate the intertype (IT) domain in strongly disordered ferromagnetic superconductors with a Curie temperature lower than the superconducting critical temperature. In such unique materials, the coexistence of superconductivity and ferromagnetism allows for the exploration of both unconventional superconductivity and interplay between magnetism and superconductivity. The study utilizes an extended Ginzburg-Landau model for the dirty limit to calculate the boundaries of the IT domain which is characterized by a complex vortex-vortex interaction and exotic vortex configurations. The analysis reveals that the IT domain in dirty ferromagnetic superconductors is not qualitatively different from the clean case and remains similarly large, suggesting that disorder does not hinder the exploration of the rich variety of IT superconductivity in ferromagnetic superconductors. This work expands understanding of the interplay between superconductivity and magnetism in complex materials and could guide future experimental studies.
\end{abstract}

\maketitle

It is known that superconducting and ferromagnetic orderings are antagonistic. The electromagnetic mechanism, which is the generation of the screening currents by the magnetic induction of the ferromagnetic system, tends to suppress superconductivity~\cite{Ginzburg-1957}. In addition, the exchange interaction favours an alignment of the electron spins in Cooper pairs in the same direction, which destroys the conventional spin-singlet Cooper pairing~\cite{Matthias-1958}. Therefore, the coexistence of superconductivity and ferromagnetism is primarily realized in artificial superconductor/ferromagnet heterostructures~\cite{Bergeret-2005,Buzdin-2005,Baek-2014,Golovchanskiy-2019,Stolyarov-2022}. In bulk systems superconductivity in most cases is either destroyed by the emergence of ferromagnetism~\cite{Moncton-1980,Lynn-1981,Zeng-2022} or coexists with non-ferromagnetic ordering such as the antiferromagnetic one~\cite{Maple-2013,Stolyarov-2020,Kim-2021}. The well-known examples of superconductivity surviving in the ferromagnetic bulk state are uranium-based superconductors~\cite{Aoki-2012}, where the pairing is spin-triplet, and thus, the superconducting ordering is not suppressed by the exchange interaction. In these compounds the Curie temperature $T_m$ is significantly higher than the superconducting critical temperature $T_c$ and the ferromagnetic ordering is dominant and just marginally influenced by the superconductivity. 

However, the coexistence of the singlet superconductivity and the ferromagnetic ordering in a wide temperature range has been recently observed in P-doped EuFe$_2$As$_2$ compound~\cite{Ren-2009,Cao-2011,Jeevan-2011,Stolyarov-2018,Grebenchuk-2020}. Interaction with the magnetic subsystem in this material is primarily electromagnetic and the ferromagnetism is weak enough to not suppress the singlet pairing. As a result, $T_m$ is less than $T_c$ for a large interval of P-doping (for example, $T_m=19$ K while $T_c=24.2$ K). Interplay between ferromagnetism and superconductivity in this compound at $T < T_m$ results in the formation of unconventional magnetic and superconducting spatial configurations~\cite{Stolyarov-2018, Grebenchuk-2020,Devizorova-2019,Vinnikov-2019} such as unusually narrow magnetic domains, Meissner domains and spontaneously generated vortex-antivortex pairs embedded in striped and dendritic patterns. 

However, the magnetic subsystem could influence superconductivity even at temperature $T>T_m$ in spite of the absence of the intrinsic magnetic ordering. It has been theoretically shown and experimentally observed in ErRh$_4$B$_4$ that strong paramagnetic response of the magnetic subsystem just above $T_m$ leads to the crossover from type-II to type-I superconductivity when the temperature is lowered down to $T_m$~\cite{Bulaevskii-1985}. The crossover is associated with the change of the magnetic penetration depth $\lambda$. $\lambda$ defines the radius of the vortex area $S$ the magnetic flux $\int \textbf{B}\cdot d\textbf{S}$ through which is equal to the superconductive ﬂux quantum $\Phi_0$. The magnetic induction $\textbf{B}$ is proportional to the magnetic permeability $\mu$, and given that $\Phi_0$ is constant, $S$ is therefore inversely proportional to $\mu$. Thus $\lambda$ is multiplied by the factor $1/\sqrt{\mu}$ and, consequently, the Ginzburg-Landau (GL) parameter $\kappa=\lambda/\xi$ is multiplied by the same factor. Since $\mu$ of the magnetic subsystem increases with a decrease in the temperature and diverges at $T_m$, the GL parameter $\kappa$ significantly decreases when the temperature is lowered to $T_m$. As a result, the initially (in the vicinity of $T_c$) type-II superconductor turns into a type-I superconductor.  

According to the GL theory, the crossover between the superconductivity types occurs at the point where $\kappa=\kappa_0=1/\sqrt{2}$, which is also known as the Bogomolnyi point (B-point)~\cite{de_Gennes,Bogomolnyi-1976}.  At the B-point vortices do not interact and the superconducting state is infinitely degenerate, which means that positions and configurations of vortices can be arbitrary. But this picture is correct only in the nearest vicinity of $T_c$. The B-point degeneracy is removed by nonlocal interactions that are not captured in the GL theory and come into play below $T_c$. Therefore, the B-point unfolds into a temperature-dependent finite domain between types I and II, which is called the intertype (IT) domain or is referred to as the domain of type-II/1 superconductivity~\cite{Jacobs-1971,Klein-1987,Miranovic-2003,Vagov-2016}. In the IT domain vortex-vortex interaction is more complex than in conventional types, it becomes spatially nonmonotonic or even has the many-body character~\cite{Brandt-2011,Silva-2015,Wolf-2017}. This interaction gives rise to exotic vortex configurations such as vortex clusters and vortex liquid droplets, leading to the magnetic response with characteristics of both the conventional superconductor types~\cite{Krageloh-1969,Essmann-1971,Weber-1989,Reimann-2017,Backs-2019,Vagov-2020,Brems-2022,Vagov-2023}. Different vortex configurations appear to be stable (or energetically favorable) in different parts of the IT domain, which defines its internal structure.

In conventional superconductors the IT domain is narrow and thus, the number of the IT superconductors among the conventional materials is limited. However, the IT domain is enlarged by including additional nonlocal interactions. For example, in sufficiently thin superconducting films the stray magnetic field outside the superconducting sample plays a role of such interactions~\cite{Cordoba-Camacho-2016,Cordoba-Camacho-2021}. The interband coupling between different band condensates~\cite{Vagov-2016} provides especially significant enlargement of the IT domain in multiband superconductors as well as a significant increase of the critical temperature~\cite{Salasnich2019,Saraiva2020,Saraiva2021}. However, even if the IT domain is broadened, its whole internal structure (and the corresponding variety of vortex configurations) can not be accessible in a given material since the GL parameter $\kappa$ is ﬁxed by microscopic parameters of the corresponding material. In this case, an initially low-kappa type-I or type-II superconductor can only be converted into an IT one when lowering the temperature. The GL parameter $\kappa$ of a material can be changed by doping the material with a nonmagnetic disorder, which simultaneously reduces the IT domain~\cite{Jacobs-1971,Kramer-1974}.

In contrast, in ferromagnetic superconductors with $T_m<T_c$, like P-doped EuFe$_2$As$_2$, it is expected that the value of $\kappa$ changes with temperature which would allow to drive the type-II superconductor into the type-I regime crossing the whole IT domain by simply lowering the temperature. Such temperature dependent crossover has been unveiled in the recent work~\cite{Vagov-2023-cp} which demonstrated that the IT domain is accessible even at $\kappa\gg\kappa_0$ by virtue of the interaction with the magnetic order parameter. The analysis in Ref.~\cite{Vagov-2023-cp} is only for the clean limit.

However, doping in the ferromagnetic superconductor (e.g., P-doped EuFe$_2$As$_2$) increases disorder, which may reduce and even eliminate the IT domain. In the present work we investigate the IT domain in strongly disordered ferromagnetic superconductors with $T_m<T_c$. For this purpose we derive the extended GL model for the dirty limit, based on which the boundaries of the IT domain are studied. A fundamental conclusion of the work is that the IT domain in dirty ferromagnetic superconductors is not qualitatively different from that observed in clean materials. Therefore, disorder does not hinder the study of the temperature-dependent IT domain.

We consider dirty ferromagnetic superconductors by combining the Usadel theory~\cite{Usadel-1970} with a phenomenological ferromagnet model. Both the superconductive and magnetic subsystems are assumed isotropic. The free energy density ($\fbb$) of the system has three contributions
\begin{equation}
\label{free_energy}
\fbb =\fbb_{\rm s} + \fbb_{\rm m} + \fbb_{\rm int},    
\end{equation}
where the superconductor contribution $\fbb_{\rm s}$ is given by
\begin{align}
\fbb_{\rm s}= \frac{{\bf B}^2}{8\pi}+ 2\pi T N(0)\sum_{\omega>0}\Big\{ 2\hbar\omega (1-g)-2\mathrm{Re}(f^*\Delta)+\frac{\hbar\cal{D}}{2}\big[|{\bf D} f|^2+(\boldsymbol{\nabla} g)^2\big]\Big\}
\label{f_s}
\end{align}
with $N(0)$ being the density of states at the Fermi level. $g=g({\bf r},\omega)$ and $f=f({\bf r},\omega)$ are the normal and anomalous quasi-classic Green functions which satisfy the normalization condition $g^2 +|f|^2=1$. $\omega$, $\Delta$ and ${\cal D}$ are the fermionic Matsubara frequency, the gap function and the diffusion coefficient, respectively. And, ${\bf D}=\boldsymbol{\nabla}-(i2e/\hbar \cbb){\bf A}$ is the gauge-invariant derivative with $e$ being the electron charge. 

The second contribution $f_{\rm m}$ is from the ferromagnetic subsystem
\begin{align}
\fbb_{\rm m}=\frac{a_{\rm m}}{2} {\bf M}^2 + \frac{b_{\rm m}}{4}({\bf M}^2)^2 + \frac{\mathcal{K}_{m}}{2} \sum_{i} (\nabla_i  {\bf M} )^2.
\label{f_m}
\end{align}
Here, ${\bf M}$ is a three-component magnetization vector, while $a_{\rm m}$, $b_{\rm m}$, and $\mathcal{K}_{\rm m}$ are the relevant coefficients. At last, the third contribution $\fbb_{\rm int}$ describes the interaction between these two subsystems: 
\begin{equation}
\fbb_{\rm int}=\gamma {\bf M}^2|\Delta|^2 - {\bf M} \cdot {\bf B}
\label{f_int}
\end{equation}
with $\gamma$ being the coupling constant. The ferromagnetic transition is governed by the temperature dependence of $a_{\rm m}=\alpha_{\rm m} (T-\theta)$ where $\theta$ is the bare Curie temperature. The minimizing of $\fbb$ in Eq. (\ref{free_energy}) determines the stable spatial $\Delta$, ${\bf B}$ and ${\bf M}$.

To study the IT domain and calculate its boundaries, we employ the perturbation approach in Refs.~\cite{Jacobs-1971,Shanenko-2011,Vagov-2012}, and keep one order beyond the GL theory with $\tau = 1- T/T_{\rm c}$ being the perturbation parameter. The derivation of this expansion is similar to the case of clean superconductors. The Green functions and the free energy density are represented as series in powers of the gap function and its spatial derivatives. Only the leading terms are kept in the series. After these steps (see the Supporting information for details), we obtain the superconducting contribution as follows:
\begin{align}
\fbb_{\rm s}= &\frac{{\bf B}^2}{8\pi}+\left[-N(0)A +a\left(\tau+\frac{\tau^2}{2}\right)\right] |\Delta|^2+\frac{b}{2}(1+2\tau)|\Delta|^4-\frac{c|\Delta|^6}{3}\notag\\
&+\mathcal{K}(1+\tau)|{\bf D}\Delta|^2 
-\mathcal{Q}|{\bf D}^2\Delta|^2-\frac{\mathcal{L}}{2}\Big\{6|\Delta|^2|{\bf D}\Delta|^2+\big[\Delta^2({\bf D}^*\Delta^*)^2+ {\rm c.c.}\big]\Big\},
\label{f_str}
\end{align}
where the coefficients are given by
\begin{align}
\nonumber
&a=- N(0), A=\ln\frac{2e^{\Gamma}\hbar\omega_D}{\pi T_c}, b=N(0)\frac{7\zeta (3)}{8\pi^2T_c^2},\notag\\ 
&\mathcal{K}=N(0)\frac{\pi\hbar\mathcal{D}}{8T_c}, \mathcal{Q}=\frac{(\hbar\mathcal{D})^2}{2} b, \mathcal{L}=N(0)\frac{\pi\hbar\mathcal{D}}{192T_c^3}, 
 c=N(0)\frac{93\zeta(5)}{128\pi^4T_c^4},
\label{coeffs}
\end{align}
with $\omega_D$ being the Debye frequency, $\zeta(x)$ the Riemann zeta function, and $\Gamma=0.577$.

It is instructive to compare the free energy density in Eq.(~\ref{f_str}) with that obtained earlier in the clean limit~\cite{Vagov-2016,Vagov-2012}. The definition of the kinetic coefficients $\mathcal{K},\mathcal{Q}$, and $\mathcal{L}$ are now given by the diffusion coefficient $\mathcal{D}$. Besides that, three terms in the expansion for the clean case that correspond to the coefficient $\mathcal{Q}$ are reduced to the single term in Eq.~(\ref{coeffs}) in the dirty limit, while the contributions proportional to $\boldsymbol{\nabla}\times{\bf B}$ and ${\bf B}^2$, are absent. Finally, the first term in the braces of Eq.~(\ref{f_str}) has a numerical factor $6$ instead of $8$ in the clean-limit case. But, the structure of the free energy density is generally similar to that of the clean superconductor.

Since in our system the Curie temperature $T_{\rm m}$ is lower than the critical temperature $T_{\rm c}$, we can assume that the magnetic order parameter ${\bf M}$ is induced by the interaction with the superconducting subsystem, and ${\bf M}$ is zero when the coupling between the subsystems is absent. Under this assumption we can seek the order parameter, ﬁeld and magnetization in the form of the following $\tau$-expansion
\begin{align}
    &\Delta = \tau^{1/2} \Psi + \tau^{3/2} \psi+\dots,\, {\bf M} = \tau \Mag + \tau^2 \boldsymbol{\mathfrak{m}} + \dots,\notag \\
    &{\bf B} = \tau \vecB + \tau^2  \boldsymbol{\mathfrak{b}} + \dots,\; 
    {\bf A} = \tau^{1/2} \vecA + \tau^{3/2} \boldsymbol{\mathfrak{a}} + \dots.
\label{tau_exp}
\end{align} 
Besides that, we take into account that in the vicinity of $T_c$ the magnetic penetration depth $\lambda$ and the GL coherence length $\xi$ diverge as $\lambda, \xi \propto \tau^{-1/2}$. To take this into account we introduce the spatial scaling ${\bf r} \to \tau^{1/2}{\bf r}$ and obtain the scaling factor for the spatial derivatives as $\boldsymbol{\nabla} \to \tau^{-1/2} {\boldsymbol{\nabla}}$. We apply the original $T-$dependence form for $a_{\rm m}$, because near the ferromagnetic transition the expansion of this coefficient yields a poor approximation when only two lowest order contributions in the perturbation series in powers of $\tau$ are taken into account.

Applying the $\tau$-expansion to the free energy functional with the density given by Eq.(\ref{free_energy}) up to the order of $\tau^3$ (see the Supporting information), we derive the expansion series for its stationary point equations. The contribution of the order $\tau$ gives the equation for $T_c$. The contribution of the order $\tau^2$ yields the GL theory equations with the linear coupling to the magnetic subsystem:
\begin{subequations}
\label{eq-GL}
\begin{align}
    &(a + b |\Psi|^2) \Psi - {\cal K}{\vecD}^2 \Psi = 0,  \\
    & {\rm rot} \big [ {\vecB} - 4\uppi {\Mag} \big] = \frac{4\uppi}{\mathbbm{c}} \boldsymbol{\mathfrak{j}}, \\
    & a_{\rm m} {\Mag} = {\vecB},
\end{align}
\end{subequations}
where $\vecD=\boldsymbol{\nabla}-(i2e/\hbar \cbb)\vecA$ and the leading order contribution to the supercurrent density is given by $\boldsymbol{\mathfrak{j}} = {\cal K} \mathbbm{c}\boldsymbol{\mathfrak{i}}$ with
\begin{equation}
    \boldsymbol{\mathfrak{i}} = \frac{4e}{\hbar \mathbbm{c}} {\rm Im}[\Psi^*\vecD \Psi].
\end{equation}
Equations (\ref{eq-GL}) describe a superconductor inside a magnetic subsystem with magnetic permeability $\mu$, and are reduced to the GL standard equations in the form
\begin{align}
    (a + b |\Psi|^2) \Psi - {\cal K}{\vecD}^2 \Psi = 0,\notag\\ 
    {\rm rot} \vecB = \frac{4\pi\mu}{\mathbbm{c}} \boldsymbol{\mathfrak{j}},
\label{eq:GLmu}
\end{align}
where the magnetic permeability is defined as
\begin{align}
    \mu = \left(1- \frac{4\uppi}{a_{\rm m}}\right)^{-1}.
    \label{eq:mu}
\end{align}
It diverges at the Curie temperature $T_{\rm m}$, given as $a_{\rm m}(T_{\rm m}) = 4\uppi$. Therefore, $\theta$ is related to $T_{\rm m}$ via $\theta=T_{\rm m} - 4\uppi/\alpha_{\rm m}$. 

The order $\tau^3$ of the stationary point equations determines $\psi$, $\boldsymbol{\mathfrak{a}}$ and $\boldsymbol{\mathfrak{m}}$. However, the explicit expressions of these quantities are not needed when we only consider the leading corrections for the GL-form free energy. Thus, to calculate the free energy up to the order $\tau^3$, we just need to know the solutions to Eqs. (\ref{eq:GLmu}).

It is convenient to introduce the dimensionless quantities
\begin{align}
    \tilde {\bf r} = \frac{\bf r}{\lambda_\mu \sqrt{2}},\,\tilde \vecB = \frac{\kappa_\mu\sqrt{2}}{\mu\mathcal{H}_c} \vecB,\, 
    \tilde \vecA = \frac{\kappa_\mu}{\mu\mathcal{H}_c \lambda_\mu} \vecA,
    \tilde \Psi = \frac{\Psi}{\Psi_0},\; 
    \tilde{\fbb} = \frac{4\pi\,\fbb}{\mu\mathcal{H}^2_c},
\label{dim_var}
\end{align}
where $\Psi_0=\sqrt{-a/b}$ is the uniform solution of GL Eqs. (\ref{eq:GLmu}) and $\mathcal{H}_c=\sqrt{4\pi a^2/b \mu}$ is the GL thermodynamic critical field (up to the factor $\tau$). The eﬀective magnetic penetration depth $\lambda_\mu$ and the eﬀective GL parameter $\kappa_\mu$ are given by
\begin{align}
\lambda_\mu= \frac{\lambda}{\sqrt{\mu}}, \; \kappa_\mu = \frac{\kappa}{\sqrt{\mu}}, 
\label{eq:lam_kap}
\end{align}
where $\lambda$ and $\kappa$ are the magnetic depth and GL parameter of the isolated superconducting subsystem. Using the dimensionless quantities, we can write the GL Eqs. (\ref{eq:GLmu}) as
\begin{align}
\Psi\big(1-|\Psi|^2\big) + \frac{1}{2 \kappa_\mu^2} \vecD^2 \Psi=0,\;
{\rm rot} \vecB=\boldsymbol{\mathfrak{i}},
\label{eq:GLsc}
\end{align}
where $\vecD=\boldsymbol{\nabla} +\mathbbm{i}\vecA$ and $\boldsymbol{\mathfrak{i}} = 2{\rm Im}[\Psi\vecD^*\Psi^*]$. Here and below we utilize these dimensionless quantities without tilde. Eqs. (\ref{eq:GLsc}) satisfy the dimensionless GL functional with the density
\begin{equation}
f^{(0)} = \frac{\vecB^2}{4 \kappa_\mu^2} +\frac{1}{2 \kappa_\mu^2}|\vecD \Psi|^2 -|\Psi|^2 +\frac{1}{2}|\Psi|^4.
\label{eq:f_GL}
\end{equation}
Therefore the GL equations and the GL free energy functional of the ferromagnetic superconductor are reduced to the standard GL formalism with the effective GL parameter $\kappa_\mu$. 

In both the clean and dirty limits, the GL formalism for ferromagnetic superconductors shows that the crossover between type I and type II occurs when $\kappa_\mu$ crosses $\kappa_0$. From Eqs. (\ref{eq:mu}) and (\ref{eq:lam_kap}) we can define the crossover line $\kappa^\ast(T)$ on the $\kappa$-$T$ plane that separates type I and type II as
\begin{align}
    \kappa^*= \kappa_0 \sqrt{\frac{T-\theta}{T-T_{\rm m}}}.
\label{eq:kappa*}
\end{align}
It can be seen that if a ferromagnetic superconductor are in type II near $T_c$ [$\kappa > \kappa^*(T_{\rm c})$] then it should cross the line separating type-I and II superconductivity when decreasing $T$. In Ref.~\cite{Vagov-2023-cp} the crossover between the two types is discussed in more detail in the framework of the GL theory. Here we remark that the temperature $T^*$, at which the crossover occurs, is determined only by the Curie temperature $T_m$ and the parameter $\alpha_{\rm m} T_{\rm c}$.

To find the boundaries of the IT domain, we compare the Gibbs free energy of the Meissner state at the thermodynamic critical field $H_c$ with that of a specific field-condensate configuration~\cite{Vagov-2016}. If the difference in Gibbs free energy between the specific configuration and the Meissner state is positive then the system is in type I. Otherwise, it is type II. The difference between the Gibbs free energies of a nonuniform field-condensate configuration and the Meissner state writes as
\begin{align}
G = \int \gbb\, d^2{\bf r} , \quad \gbb = \fbb +\frac{H^2_{\rm c}}{2}- \frac{H_{\rm c} B}{\sqrt{2}\kappa_\mu},
\label{Gibbs_energy}
\end{align}
where we assume that the external field ${\bf H}_{\rm c} = (0,0,H_{\rm c})$ is parallel to ${\bf B} = (0,0,B)$, and, $\gbb$ and $G$ are given in units of $\mu \mathcal{H}_{\rm c}^2/4\uppi$ and $\mu \mathcal{H}_{\rm c}^2 \lambda_\mu^2 L/2\uppi$, respectively. $L$ is the sample size in the $z$-direction. The energy difference is calculated by using relevant solutions for the stationary point Eqs. (\ref{eq:GLsc}) which are independent of $z$, therefore, the integration is performed in the $x$-$y$ plane.

Using the $\tau$-expansion approach, we represent $G$ as a series in $\tau$, and keep only the leading correction to the GL contribution (see the Supporting information). Besides that, we employ the expansion in the small deviation $\delta\kappa_\mu = \kappa_\mu - \kappa_0$, since we are interested in the IT domain near $\kappa_0$. Then the expansion yields
\begin{align}
G = \tau^2\Big(G^{(0)} + \frac{d G^{(0)}}{d\kappa_\mu}\,\delta\kappa_\mu + 
G^{(1)} \,\tau\Big),
\label{eq:Gibbs_expansion}
\end{align}
where $G^{(0)}$ is the GL Gibbs free energy at $\kappa_0$, $dG^{(0)}/d\kappa_\mu$ is its derivative with respect to $\kappa_\mu$ and $G^{(1)}$ is the leading order correction in $\tau$ expansion also calculated at $\kappa_0$. At $\kappa_0$ the GL theory is reduced to the ﬁrst-order Bogomolnyi equations 
\begin{align}
\mathcal{D}_{-}\Psi=0, \;\mathcal{B} = 1 - |\Psi|^2, 
\label{eq:Bog}
\end{align}
where $\mathcal{D}_{-}=\mathcal{D}_x - \mathbbm{i}\mathcal{D}_y$, with $\mathcal{D}_x, \mathcal{D}_y$ the $x$ and $y$ components of $\vecD$. Using these equations, we obtain the following expression for the Gibbs free-energy difference
\begin{align}
\frac{G}{\tau^2}= - \sqrt{2}\,  {\cal I} \, \delta \kappa_\mu + \tau \, \Big\{\big[{\cal Q}-c + \gamma \big] \,{\cal I}  + \Big[\frac{3{\cal L}}{2}-  c-{\cal Q} -\gamma  + \mu {\cal K}_{\rm m}\Big] {\cal J} \Big\},
\label{eq:functional_exp_final}
\end{align}
where the dimensionless coefficients are defined as
\begin{align}
&c=\frac{c a }{3b^2}, \quad  {\cal Q} = \frac{ {\cal Q}
a}{{\cal K}^2}, \quad {\cal L}=\frac{{\cal L}\,a}{b\, {\cal
K}}, \quad {\cal K}_{m}  =-\frac{2\pi a {\cal K}_m}{a^2_m{\cal K} }, \quad  \gamma = -\frac{\gamma a \Phi^2_0}{4\uppi^2 a_{\rm m}^2 {\cal K}^2}, 
\label{eq:constants}
\end{align}
with
\begin{align}
{\cal I} =\!\! \int\! |\Psi|^2 \big(1 - |\Psi|^2\big)d^2{\bf r},\, {\cal J} =\!\! \int\! |\Psi|^4 \big(1 - |\Psi|^2\big)d^2{\bf r}.
\label{eq:I_J}
\end{align}
Here $\Psi$ is a solution of the Bogomolnyi equations for a particular condensate configuration. 

In contrast to the Gibbs free-energy difference obtained in Refs.~\cite{Vagov-2016,Vagov-2020}, the small quantities $\delta\kappa_\mu$ and $\tau$ are dependent because $\kappa_\mu$ varies with $T$. But the form of the Gibbs free-energy difference in Eq. (\ref{eq:functional_exp_final}) for ferromagnetic superconductors is the same as that of conventional superconductors, which shows that  the criteria of identifying the IT domain boundary in the $\kappa$-$T$ plane for both conventional and ferromagnetic superconductors are also the same.~\cite{Vagov-2016,Wolf-2017,Vagov-2020} From the condition $G=0$ we find the expression for the critical GL parameter $\kappa_c$ which separates domains with and without the particular condensate configuration in the $\kappa$-$T$ plane
\begin{align}
    \frac{\kappa_c}{\kappa_0  \sqrt{\mu} } =   1 +  \tau \left[{\cal Q}-c + \gamma+\frac{\cal J}{\cal I}\left(\frac{3{\cal L}}{2}-  c-{\cal Q} -\gamma  + \mu {\cal K}_{\rm m}\right) \right] .
\label{kappa-crit}
\end{align}

The boundary of the IT domain can be determined by two criteria. The first criterion is the condition of the appearance/disappearance of a nonuniform superconducting state for the fields above $H_c$ which means the equality of the thermodynamic and upper critical fields $H_{\rm c} = H_{{\rm c}2}$. In this case $\Psi\to 0$, so we can use ${\cal J}/{\cal I}\to 0$, and obtain
\begin{align}
    \frac{\kappa_1}{\kappa_0  \sqrt{\mu} } =   1 +  \tau \left({\cal Q}-c + \gamma\right) .
\label{kappa-1}
\end{align}

The second criterion is defined by the onset of the long-range vortex-vortex attraction. So we need to find the asymptote of the two-vortex solution of the GL equations at large distance $R$ between vortices which gives ${\cal J}/{\cal I}=2$~\cite{Vagov-2016}. Using this result, we get
\begin{align}
    \frac{\kappa_2}{\kappa_0  \sqrt{\mu} } =   1 +  \tau \left(-{\cal Q}-3c +3{\cal L}- \gamma+2\mu {\cal K}_{\rm m}\right) .
\label{kappa-2}
\end{align}
Using Eqs. (\ref{coeffs}), it can be shown that the dimensionless coefficients of the superconducting subsystem are independent of microscopic parameters:
\begin{align}
c = -0.227, \quad {\cal L} = -0.391, \quad  {\cal Q} = -0.3455.
\end{align}
Notice that the constants ${\cal L}$ and ${\cal Q}$ differ from the clean case. The other coefficients depend on the microscopic characteristics of both the superconducting and magnetic subsystems. 

\begin{figure*}
\centering
\includegraphics[width=1.0\linewidth]{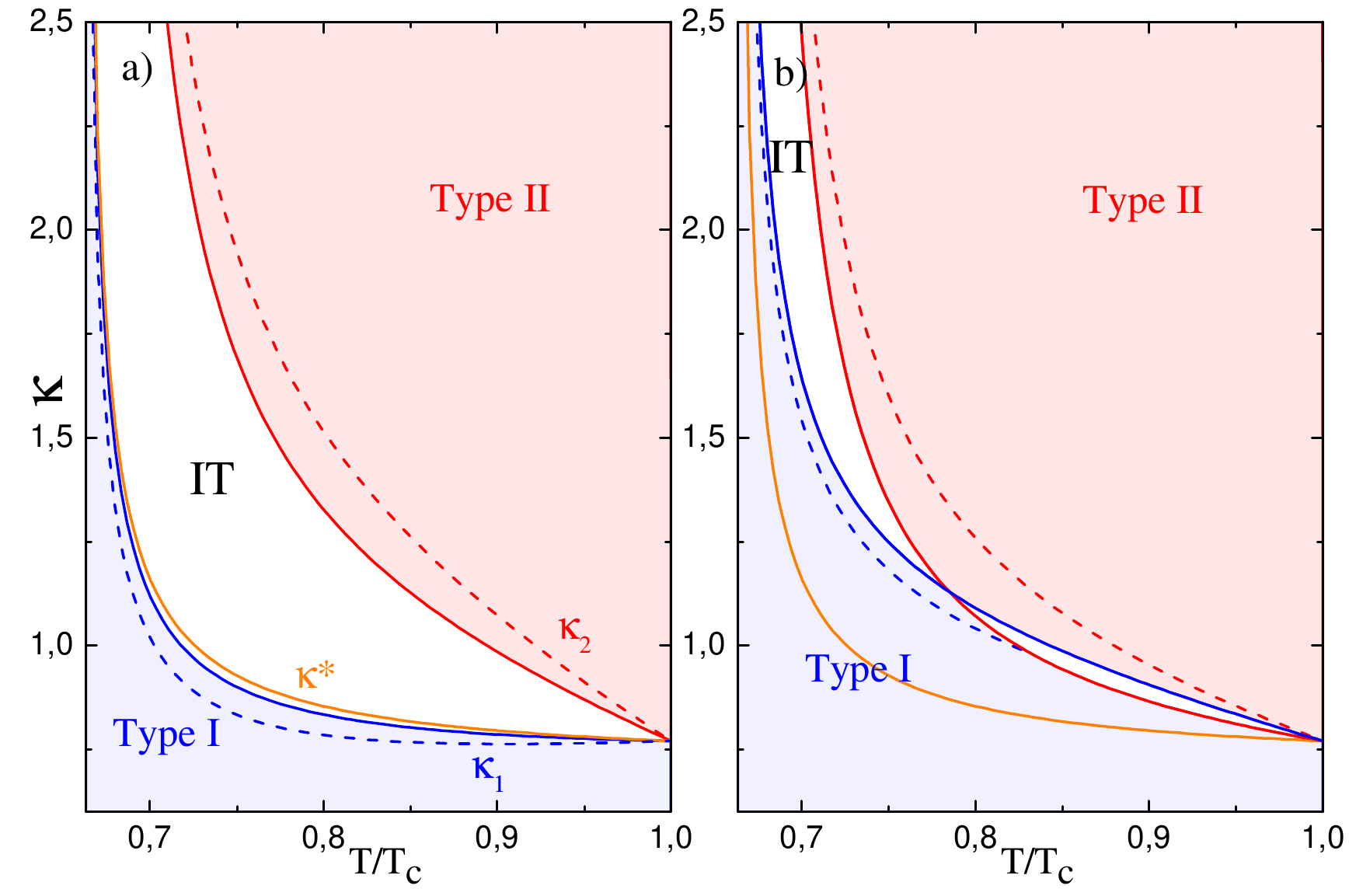}
\caption{The $\kappa$-$T$ phase diagram for a ferromagnetic superconductor with the Curie temperature smaller than the superconducting one, where $\kappa$ is the Ginzburg-Landau parameter and $T$ is the temperature. Panels show the boundaries $\kappa_1$ and $\kappa_2$ of the intertype (IT) domain for the coupling constant $\gamma=0$ (a) and $\gamma=1.5$ (b). Solid lines correspond to the dirty ferromagnetic superconductor and dashed lines correspond to the clean one. $\kappa^\ast$ separates types I and II in the GL theory. } 
\label{fig1}
\end{figure*}

In dirty conventional superconductors $\kappa_1$ defines the upper boundary of the IT domain and $\kappa_2$ determines the lower boundary (in contrast to clean conventional superconductors)~\cite{Jacobs-1971,Jacobs-prb}. But the situation changes for dirty ferromagnetic superconductors. For illustration we choose the same dimensionless parameters as in Ref.~\cite{Vagov-2023-cp}, i.e., we take $\gamma = 0$, ${\cal K}_{\rm m} = 1$, $\theta/T_{\rm c} =0.6$ and $\alpha_{\rm m} T_{\rm c} = 200$. It gives the Curie temperature $T_{\rm m}/T_{\rm c} = 0.66$. Using these parameters, we calculate the boundaries of the IT domain and compare them with the boundaries in the clean case. The results are shown in Fig.~\ref{fig1}(a). It can be seen that the IT domain is not only greatly expanded in comparison with a dirty conventional superconducting system but it also acquires the structure akin to that of clean superconductors. This is mainly due to the term $\mu {\cal K}_{\rm m}$ which shifts up the boundary $\kappa_2 (T)$. In comparison with the clean ferromagnetic superconductor the IT domain shrinks mildly but it still remains pretty large even at relatively large $\kappa$. The IT domain width changes weakly because in both cases it is mostly determined by the term $\mu {\cal K}_{\rm m}$. Since the IT domain has the same structure as in the clean limit, the transformation of the vortex matter with decreasing temperature is the same too. Thus, at $\kappa_2$, the interaction between two vortices is repulsive (attractive) when their distance is sufficiently short (long), which results into the formation of vortex clusters. With further decrease in temperature, the pairwise vortex interaction becomes fully attractive while the many-vortex interactions start to play a significant role, so we obtain vortex liquid droplets.

However, the coupling constant $\gamma$ reduces the size of the IT domain, and can even completely cancel the contribution of the term $\mu {\cal K}_{\rm m}$ and reverse the order of $\kappa_c$ of a dirty ferromagnetic superconductor back to the case of a dirty superconductor [see Fig.~\ref{fig1}(b)]. It happens at strong coupling, more specifically, when the coupling constant $\gamma>\mu {\cal K}_{\rm m}+3{\cal L}/2-c-{\cal Q}$. Since lowering the temperature increases $\mu$, at some temperature there is an intersection of different boundaries which separates the IT domain into two parts. We have $\kappa_1<\kappa_2$ and an IT domain just like that in a clean ferromagnetic superconductor when $T$ is lower than the intersection temperature. For $T$ higher than the intersection temperature, one gets $\kappa_2<\kappa_1$ and an IT domain similar to the case in a conventional dirty superconductor. The peculiarities of vortex interactions and configurations in the IT domain of the conventional dirty superconductor will be discussed elsewhere. Here we remark that since $\kappa>\kappa_2$ the interaction between vortices is always repulsive at large distances in such IT domain.

Concluding, we have investigated the IT domain in dirty ferromagnetic superconductors with the Curie temperature smaller than the superconducting critical temperature. In such system the interaction between superconducting and magnetic subsystems results in a temperature dependent crossover from type II to type I above the Curie temperature. Deriving and using the perturbation theory up to the leading correction to the GL theory, we have shown that the nonmagnetic disorder only weakly affects the IT domain unless the coupling between the superconducting and magnetic subsystems is sufficiently strong. In the latter case there is a temperature at which the internal structure of the IT domain changes. Thus, similarly to the clean system investigated in Ref.~\cite{Vagov-2023-cp}, a dirty ferromagnetic superconductor is a fascinating testing ground to probe details of the IT regime and its exotic field-condensate configurations. Based on our results, we expect that a moderate disorder in ferromagnetic superconductors has no significant effect on the IT domain.

\section*{Supporting Information }
Supporting Information: Theoretical derivation of the major analytical results.

\section*{Acknowledgements}

The work was financed within the framework of the Basic Research Program of HSE University and Science Foundation of Zhejiang Sci-Tech University(ZSTU) (Grants No. 19062463-Y).

\end{document}